\documentclass[11pt]{revtex4}
\begin{document}
\newcommand{\dfrac}[2]{\frac{\displaystyle #1}{\displaystyle #2}}
\baselineskip=14pt
\title{Difermion condensates in vacuum in 2-4D
four-fermion interaction models\footnote{The project supported by
the National Natural Science Foundation of China under Grant
No.10475113.}}
\author{
 Bang-Rong Zhou\footnote{Electronic mailing address:
zhoubr@163bj.com}}  \affiliation{College of Physical Sciences,
Graduate School of \\ the Chinese Academy of Sciences, Beijing
100049, China}

 \vspace{2cm}
\begin{abstract}
In any four fermion (denoted by $q$) interaction models,
   the couplings of $(qq)^2$-form  can always coexist
   with the ones of $(\bar{q} q)^2$-form via the Fierz
   transformations. Hence, even in vacuum, there could be
    interplay between the condensates $\langle\bar{q} q\rangle$ and $\langle qq\rangle$.
    Theoretical analysis of this problem is generally made by
    relativistic effective potentials in the mean field approximation
     in 2D, 3D and 4D models with two flavor and $N_c$ color massless
      fermions. \\
      \indent It is found that in ground states of these models,
      interplay between the two condensates mainly depend on the ratio
       $G_S/H_S$ for 2D and 4D case or $G_S/H_P$ for 3D case, where $G_S$, $H_S$ and
       $H_P$ are respectively the coupling constants in a scalar $(\bar{q}q)$,
       a scalar $(qq)$ and a pseudoscalar $(qq)$ channel. \\ \indent In ground states of
        all the models, only pure $\langle\bar{q}q\rangle$ condensates could exist if
        $G_S/H_S$ or $G_S/H_P$ is bigger than the critical value $2/N_c$,
        the ratio of the color numbers of the fermions entering into
        the condensates $\langle qq\rangle$ and $\langle\bar{q}q\rangle$.  Below it,
        differences of the models will manifest themselves. \\
        \indent In the 4D Nambu-Jona-Lasinio (NJL) model, as $G_S/H_S$ decreases to the
        region below $2/N_c$, one will first have a coexistence phase of the two
        condensates then a pure $\langle qq\rangle$ condensate phase.  Similar results
        come from a renormalized effective potential in the 2D Gross-Neveu model,
        except that the pure $\langle qq\rangle$
       condensates could exist only if $G_S/H_S=0$. In a 3D Gross-Neveu model,
       when $G_S/H_P < 2/N_c$, the phase transition similar to the 4D case can
       arise only if $N_c>4$, and for smaller $N_c$, only a pure $\langle qq\rangle$
       condensate phase exists but no coexistence phase of the two condensates happens.
       The $G_S-H_S$ (or $G_S-H_P$) phase diagrams in these models are given. \\
       \indent The results deepen our understanding of dynamical phase
       structure of four-fermion interaction models in vacuum. In addition, in view of
       absence of difermion condensates in vacuum of QCD, they will also imply a real
       restriction to any given two-flavor QCD-analogous NJL model, i.e. in the model,
       the derived smallest ratio $G_S/H_S$ via the Fierz transformations in the
       Hartree approximation must be bigger than 2/3.
\end{abstract}
\maketitle
\newpage \section{Main results}
 We have researched interplay between the fermion($q$)-antifermion
($\bar{q}$) condensates $\langle\bar{q}q\rangle$ and the difermion
condensates $\langle qq\rangle$ in vacuum in 2D, 3D and 4D
four-fermion interaction models with two flavor and $N_c$ color
massless fermions. It is found that the ground states of the
systems could be in different phases shown in the following
$G_S-H_S$ and $G_S-H_P$ phase diagrams [Fig.(a)--Fig.(d)],  where
\begin{description}
    \item[$G_S$] --- coupling constant of scalar $(\bar{q}q)^2$ channel
    \item[$H_S$] --- coupling constant of scalar color $\dfrac{N_c(N_c-1)}{2}\;-$plet
    $(qq)^2$ channel
     (4D, 2D case)
    \item[$H_P$] --- coupling constant of pseudoscalar color $\dfrac{N_c(N_c-1)}{2}\;-$plet $(qq)^2$ channel
     (3D case)
    \item[$\Lambda$] --- Euclidean Momentum cutoff of loop integrals (4D,3D case)
    \item $(\sigma_1,0)$ ---  pure $\langle\bar{q}q\rangle$ phase
    \item $(0,\Delta_1)$ ---   pure $\langle qq\rangle$ phase
    \item $(\sigma_2,\Delta_2)$ ---  mixed phase with both $\langle \bar{q}q\rangle$  and
                                     $\langle qq\rangle$
\end{description}
\vspace{3.5cm} \centerline{\Large Fig.(a)-Fig.(d) \  (pages
3-6)}\vspace{3.5cm}
\end{document}